\DocumentMetadata{}
\documentclass[sigconf,nonacm]{acmart}

\usepackage[yyyymmdd]{datetime}

\usepackage{tabularx}
\usepackage{dcolumn} 
\newcolumntype{d}[1]{D{.}{.}{#1}}

\usepackage{subcaption}

\DeclareRobustCommand{\IVIllusion}{$STIM$}
\DeclareRobustCommand{\IVTemp}{${T}_{Phys}$}

\DeclareRobustCommand{\lvlSteam}{$Steam$}
\DeclareRobustCommand{\lvlSizzle}{$Sizzle~ Sound$}
\DeclareRobustCommand{\lvlIceCubes}{$Ice~~Cubes$}
\DeclareRobustCommand{\lvlIceCubesSound}{$Ice~Cubes~Sound$}

\DeclareRobustCommand{\baseline}{$Baseline$}
\DeclareRobustCommand{\warmV}{${Warm}_{V}$}
\DeclareRobustCommand{\warmA}{${Warm}_{A}$}
\DeclareRobustCommand{\warmVA}{${Warm}_{VA}$}
\DeclareRobustCommand{\coldV}{${Cold}_{V}$}
\DeclareRobustCommand{\coldA}{${Cold}_{A}$}
\DeclareRobustCommand{\coldVA}{${Cold}_{VA}$}

\DeclareRobustCommand{\celsius}{\textdegree{}C}

\DeclareRobustCommand{\RefApparatus}[1]{\hyperref[fig:apparatus]{Figure~\ref*{fig:apparatus}#1}}

\AtBeginDocument{%
  }

\setcopyright{acmlicensed}
\copyrightyear{2025}
\acmYear{2025}
\acmDOI{XXXXXXX.XXXXXXX}
\acmConference[XX '25]{}

\begin{document}

\title{Quantifying the Effect of Thermal Illusions in Virtual Reality}


\settopmatter{authorsperrow=4}

\author{Yannick Weiss}
\orcid{0000-0003-1125-7963}
\affiliation{%
  \institution{LMU Munich}
  \city{Munich}
  \country{Germany}
}
\email{yannick.weiss@ifi.lmu.de}

\author{Marlene Eder}
\affiliation{%
  \institution{LMU Munich}
  \city{Munich}
  \postcode{80337}
  \country{Germany}}
\email{m.eder@campus.lmu.de}

\author{Oguzhan Cesur}
\affiliation{%
  \institution{LMU Munich}
  \city{Munich}
  \postcode{80337}
  \country{Germany}}
\email{o.cesur@campus.lmu.de}

\author{Steeven Villa}
\orcid{0000-0002-4881-1350}
\affiliation{%
  \institution{LMU Munich}
  \city{Munich}
  \postcode{80337}
  \country{Germany}}
\email{villa@posthci.com}

\renewcommand{\shortauthors}{Weiss et al.}

\begin{abstract}
Thermal sensations are central to how we experience the world, yet most virtual and extended reality systems fail to simulate them effectively. While hardware-based thermal displays can provide accurate temperature changes, they are often bulky, power-intensive, and restrict user mobility. Consequently, recent works have explored thermal illusions, perceptual effects that rely on cross-modal interactions, to achieve thermal experiences without physical heating or cooling. While thermal illusions have been shown to consistently alter subjective ratings, the actual extent of their effect on the perceived temperature of interacted objects remains unexplored. To address this, we contribute the findings of two user studies following psychophysical procedures. We first ordered and scaled the effects of a variety of visual and auditory cues (N=20) and subsequently quantified their isolated and combined efficacy in offsetting physical temperature changes (N=24). We found that thermal illusions elicited robust changes in subjective judgments, and auditory cues showed potential as an alternative or complementary approach to established visual techniques. However, the actual effects induced by thermal illusions were relatively small (±0.5°C) and did not consistently align with abstract ratings, suggesting a need to reconsider how future thermal illusions or experiences are designed and evaluated.
\end{abstract}




\maketitle

\section{Introduction}
From enjoying a hot cup of tea, receiving a warm embrace from friends, to checking the bathwater before hopping in, our sense of temperature plays a crucial role in our daily lives, helping us experience and interact with our environment and each other. Although thermal sensations are an essential component of our experience of the real world, current virtual (VR) or extended reality (XR) systems generally lack the capabilities to simulate these haptic experiences. 
To address this, researchers have developed various room-scale~\cite{Shaw2019_IRHeater, Hand2018_HotAirBlower_PlusIR} and wearable~\cite{Wang2024_PeltierGloves, Chen2017_Peltier_OnHmd} thermal displays to apply temperature changes to users' skin. However, these systems require the integration of additional hardware, limiting users' mobility, massively increasing power consumption, and preventing interactions and experiences beyond the specific device. 
Consequently, researchers explored novel alternative and complementary approaches using sensory illusions to manipulate temperature sensations. By exploiting multisensory interactions between visual and thermal cues, these illusions have been shown to effectively alter the perceived temperatures of virtual objects or environments, for instance, via changes in colors~\cite{Villa2024_UltrasoundAndHueHeat, Kaethner2019_HI_RedBlue_ChangesPainIntensity}, displaying fire or ice metaphors~\cite{Guenther2020_Hydraulic_PlusHIVisual, Blaga2020_HI_VisualPresentations}, or changing virtual surroundings~\cite{Brooks2020_TrigeminalNerve, Kocur2023_BurningHandsChangesBodyTemp}.

However, the effectiveness of these approaches is always measured and compared using subjective, abstract scales~\cite{Kaethner2019_HI_RedBlue_ChangesPainIntensity, Guenther2020_Hydraulic_PlusHIVisual, Peiris2017_PeltierInHMD, Villa2024_UltrasoundAndHueHeat}. The actual perceptual shift in temperature that these visuo-thermal illusions induce has not been explored. This leaves open the fundamental question of whether thermal illusions are capable of inducing large temperature alterations akin to hardware-based approaches or solely serve as tie-breakers when consciously sorting or mapping temperature sensations. 

To address this gap, we investigate and quantify the efficacy of thermal illusions. We employed two established psychophysical procedures to first order and scale, and subsequently quantify the effect of a set of thermal illusions on the perceived temperature of an interacted object. 
In addition, while prior research has primarily focused on visual cues~\cite{Blaga2020_HI_VisualPresentations, Guenther2020_Hydraulic_PlusHIVisual, Villa2024_UltrasoundAndHueHeat}, we expand this scope by examining a broader range of visual, auditory, and visuo-auditory stimuli. 
Across the two studies, we found that thermal illusions reliably influenced subjective temperature judgments, with auditory cues proving particularly effective both alone and in combination with visual input. However, we observed an incongruence between subjective ratings and actual perceptual shifts, exposing a key limitation of relying solely on abstract evaluation methods. Crucially, direct measurements revealed that the illusions produced only small changes in perceived temperature (within $\pm$0.5\celsius{}), indicating that their perceptual impact is far weaker than what prior evaluations relying on subjective scales might suggest. While these illusions did not impair perceptual sensitivity and can still serve as subtle enhancements, our findings call into question their suitability as substitutes for physical thermal feedback.

\section{Related Work}
This work draws from a large body of research aiming to understand and enhance thermal experiences in virtual environments. In the following, we provide an overview of conventional thermal rendering approaches (\autoref{RW:Devices}) and prior explorations of thermal illusions (\autoref{RW:Illusions}). 

\subsection{Devices for Thermal Rendering}
\label{RW:Devices}
Rendering realistic thermal sensations in virtual environments has been a large area of haptics research (see \cite{Park2025_ThermalWearable_Review, Ichihashi2025_ThermalDevices_Review} for recent reviews), due to its importance for immersion~\cite{Shaw2019_IRHeater}, affective communication~\cite{MotivationTemp_InterpersonalWamrth}, and comfort~\cite{Knierim2024_MotivationTemp_Comfort}.
Thermal sensations of warm and cold are primarily perceived through cutaneous receptors~\cite{Jones2008_Thermal_Basics}, leading a majority of approaches to focus on wearable devices constantly in contact with the skin~\cite{Park2025_ThermalWearable_Review}. Research primarily relies on Peltier elements placed on users' hands~\cite{Nasser2024_FingersPeltier, Wang2024_PeltierGloves}, arms~\cite{Peiris2019_ThermalBracelet, Kang2024_Peltier_Underarm}, or mounted on a head-mounted display~\cite{Peiris2017_PeltierInHMD, Chen2017_Peltier_OnHmd, Ranasinghe2018_PeltierInHMD_PlusWind, Wolf2019_PeltierInHMD_AddedVisuals}, enabling both warm and cold sensations depending on polarity. Peltiers are small and easily controlled; however, these devices require large heatsinks and power supplies to remain effective, making them generally very rigid, which restricts users' dexterity. Therefore, novel research also looked at pneumatic~\cite{Cai2020_PneumaticGlove} or liquid~\cite{Guenther2020_Hydraulic_PlusHIVisual} alternatives, which can be delivered to the skin through tubes. While these allow more dexterous interactions by being able to be bent and deformed, they still consume a large amount of power and tether the user to a large external device. Wearable approaches generally require constant contact with the skin, obstructing other haptic feedback or physical interactions. For controlled environments, research therefore applied alternate approaches, using external devices to deliver temperature feedback with encounter-type systems~\cite{Villa2024_EncounterTypeThermal} or through the air, such as infrared heaters~\cite{Shaw2019_IRHeater}, hot air blowers~\cite{Hand2018_HotAirBlower_PlusIR}, or moving heated air precisely using ultrasound~\cite{Wang2023_Midair_HeatedAirAndUltrasound, Singhal2021_MidAirHeat_PeltierAndUltrasonic}. This allows users to interact with their hands freely while still receiving thermal sensations. However, these systems require complex hardware setups that cannot easily be moved or scaled to different scenarios and environments. 

Generally, thermal rendering devices are widespread and simple to control and integrate, but severely restrict mobility, portability, and scalability due to complicated hardware setups and enormous power needs. Consequently, research increasingly explores thermal illusions, which aim to induce thermal sensations without actual physical temperature changes.

\subsection{Thermal Illusions}
\label{RW:Illusions}
A direct approach to induce the illusion of a temperature change is to stimulate the necessary receptors through chemical reactions~\cite{Lu2021_ChemicalThermal, Brooks2020_TrigeminalNerve}. However, the stimulation is rather slow, and the use of chemical reagents creates practical issues regarding storage and targeted delivery to the receptors. 

Instead, a larger area of research focuses on the use of visual cues to affect the perceived temperatures of interacted objects or environments. These illusions rely on the fact that our perception is shaped by cross-modal interactions, where the information displayed to various senses is integrated into a combined percept. This allows one sense, such as vision, to partially influence and override the sensations received from another sense, for instance, our haptic or thermal senses. This phenomenon, called multisensory integration~\cite{Ernst2004_MultisensoryIntegration}, has already been commonly investigated for visuo-haptic interactions~\cite{Ernst2002_MLE} and utilized to provide a variety of pseudo-haptic sensations in virtual environment, such as altering the perceived shape~\cite{Ban2014_HandDistortion_Shape}, size~\cite{Bergstroem2019_HandDistortion_Size}, weight~\cite{Samad2019_VHI_PseudoHapticWeight}, stiffness~\cite{Weiss2023_VHI_Stiff}, and surface texture~\cite{Etzi2018_VHI_Roughness_VisualAndAuditoryMaterials} of haptically explored objects (see \cite{IllusionsSLR} for an overview). These investigations show haptic illusions to be an effective, adaptable, and low-energy alternative to complex hardware-based rendering approaches. 

For temperature perception, a common approach builds on the hue-heat hypothesis~\cite{Villa2024_UltrasoundAndHueHeat, Kaethner2019_HI_RedBlue_ChangesPainIntensity, Blaga2020_HI_VisualPresentations, Chinazzo2021_BlueOrange}, which proposes that certain color hues are systematically associated with thermal sensations. For example, red hues tend to increase, while blue hues tend to decrease, the perceived temperature of an object~\cite{Ziat2016_HueHeat}.
\citet{Villa2024_EncounterTypeThermal} investigated representations of gaseous, liquid, and solid virtual objects haptically presented in mid-air by an ultrasound haptic display and the effect of color hues (red, blue, gray) of these objects on perceived temperature. They found that red was robustly identified as warmer than blue or gray, but that objects in the neutral gray tone were judged colder than their blue counterparts. 
Hue-heat associations have further been employed in VR to modify perceived pain intensity of hot water by visually displaying blue and red indicators in the virtual environment~\cite{Kaethner2019_HI_RedBlue_ChangesPainIntensity}. 
Beyond color adjustments, many approaches rely on a variety of explicit visualizations commonly associated with warmth, heat, or coolness. These include visualizations of fire~\cite{Guenther2020_Hydraulic_PlusHIVisual, Weir2013_BurnignHands, Shaw2019_IRHeater, Peiris2017_PeltierInHMD} and ice~\cite{Blaga2020_HI_VisualPresentations}, steam as an indirect cue of hot liquids~\cite{Blaga2020_HI_VisualPresentations}, common appliances such as heaters or furnaces~\cite{Guenther2020_Hydraulic_PlusHIVisual, Brooks2020_TrigeminalNerve}, weather effects~\cite{Guenther2020_Hydraulic_PlusHIVisual, Peiris2017_PeltierInHMD}, or hot and cold environments~\cite{Brooks2020_TrigeminalNerve, Blaga2020_HI_VisualPresentations, Kocur2023_BurningHandsChangesBodyTemp, Ranasinghe2017_PeltierOnNeck_PlusWind}. These approaches are often coupled with thermal rendering devices to enhance their combined effect~\cite{Guenther2020_Hydraulic_PlusHIVisual, Shaw2019_IRHeater, Brooks2020_TrigeminalNerve, Ranasinghe2017_PeltierOnNeck_PlusWind, Peiris2017_PeltierInHMD}.

These approaches have shown that the visuo-thermal illusions effectively change subjective judgements~\cite{Villa2024_UltrasoundAndHueHeat, Guenther2020_Hydraulic_PlusHIVisual}, the users' behaviour~\cite{Blaga2020_HI_VisualPresentations}, and even evoke physiological changes such as alterations of body temperature~\cite{Kocur2023_BurningHandsChangesBodyTemp}.
However, abstract scales~\cite{Kaethner2019_HI_RedBlue_ChangesPainIntensity, Guenther2020_Hydraulic_PlusHIVisual, Peiris2017_PeltierInHMD, Villa2024_UltrasoundAndHueHeat} or qualitative feedback~\cite{Blaga2020_HI_VisualPresentations} are unable to assess how much thermal illusions truly shift the perceived temperatures of touched virtual objects. 
Thus, it remains unclear whether thermal illusions suffice to replace thermal devices or only allow for small differentiations based on conscious choices. Further, prior approaches generally rely on the influence of visual representations, leaving a gap concerning possible auditory or multimodal approaches to alter perceived temperatures. 
In this work, we address these gaps by quantifying the size of the perceptual shift in temperatures induced by visual, auditory, and visuo-auditory cues.

\section{Study Design}
To quantify the effect of visuo-auditory thermal illusions, we conducted two controlled user studies following established psychophysical protocols~\cite{Jones2013_PsychophysicsInHaptics, Gescheider1997_Psychophysics}. We first produced a varied set of visuo-thermal and auditory-thermal illusions conveying both increases and decreases in temperature. This selection was informed by prior established approaches to thermal illusions~\cite{Ziat2016_HueHeat, Villa2024_UltrasoundAndHueHeat, Kaethner2019_HI_RedBlue_ChangesPainIntensity, Guenther2020_Hydraulic_PlusHIVisual, Blaga2020_HI_VisualPresentations, Kocur2023_BurningHandsChangesBodyTemp}. We then conducted a magnitude estimation task to rank and scale these illusions and find their relative effect on perceived temperature. After identifying robust visual and auditory cues for both warmer and colder sensations, we used a discrimination task to compare the visual, auditory, and combined visuo-auditory cues against physical temperature alterations to quantify their combined and isolated efficacy in shifting temperature perception.

\subsection{Stimuli}
\label{Stimuli}

Prior work has explored a large number of different (visuo-) thermal illusions in virtual environments. However, they are difficult to directly compare and rank due to the differences in visual quality and fidelity, the style of the respective representations, and the employed task designs and evaluation metrics. 
Consequently, we produced a range of visual cue designs derived from previous research to use in the same environment and task. We decided to target thermal illusions related to water, as this has commonly-experienced reactions to changes in temperatures (e.g., freezing, steaming, boiling) and has been prevalently used in prior work~\cite{Blaga2020_HI_VisualPresentations, Villa2024_UltrasoundAndHueHeat}.

\begin{figure}[t]
    \centering
    \includegraphics[width=\linewidth]{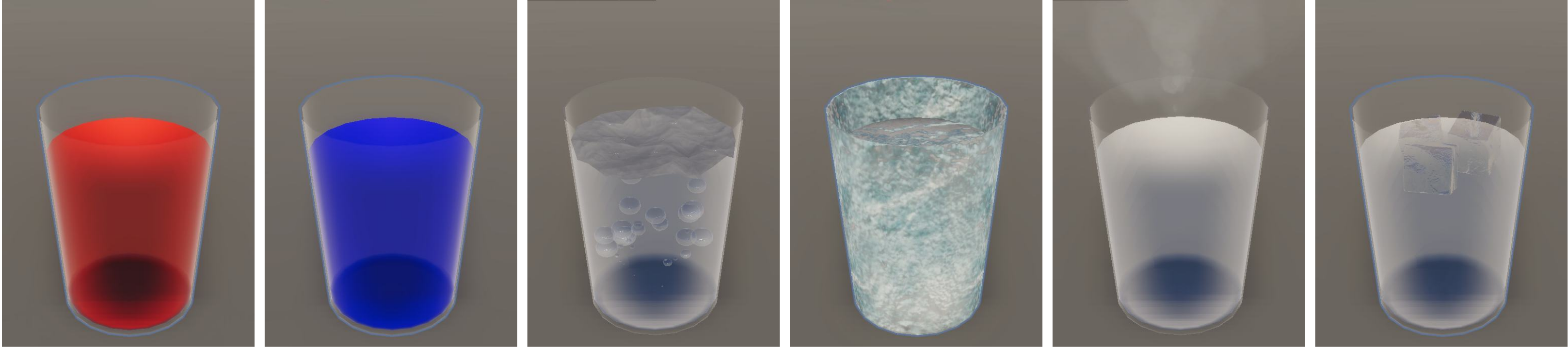}
    \caption{Investigated visual stimuli (left to right): Red \& blue coloring, boiling water, frozen glass, steam, and ice cubes.}
    \label{fig:stimuli}
    \Description{This figure shows six virtual glasses filled with liquid to represent the six visual illusions used in the study: red and blue coloring, boiling water, frozen glass, steam above the glass, and ice cubes on the liquid's surface.}
\end{figure}

In addition to a baseline, we generated a total of 10 different cue designs, comprising 6 visual and 4 auditory illusions. The visual cues are displayed in \autoref{fig:stimuli}. Auditory stimuli can be found in our repository (see \autoref{OpenScience}). The 10 visual or auditory designs correspond to different dimensions of presentation, each with a pair of cues for targeting an increased and a decreased temperature perception, respectively:

\begin{description}
    \item[Baseline] The baseline condition shows a transparent cylindrical glass filled with water. To limit the potential confounding effects of color and environment, we present the water as a semi-transparent liquid over a gray background scene. 
    \item[Visual - Hue Change] In line with prior works investigating Hue-Heat associations~\cite{Ziat2016_HueHeat, Villa2024_UltrasoundAndHueHeat, Kaethner2019_HI_RedBlue_ChangesPainIntensity}, we presented the liquid inside the our glass as either $Red$ or $Blue$. Based on prior findings, we expect these to induce a warmer and colder sensation, respectively.
    \item[Visual - Direct Cues] For direct cues, we selected visualizations that directly alter the presented state of the liquid inside the glass. For this, we generated an animated $Boiling$ condition (expected warmer) and a $Frozen$ condition (expected colder, cf.~\cite{Kocur2023_BurningHandsChangesBodyTemp}).
    \item[Visual - Indirect Cues] For indirect cues, we produced cues that do not directly alter the liquid presentation but give contextual clues about the water's temperature. For this, we render $Steam$ above the surface of the water (expected warmer, cf.~\cite{Blaga2020_HI_VisualPresentations}) or three $Ice~Cubes$ swimming on the water's surface (expected colder, cf.~\cite{Blaga2020_HI_VisualPresentations}).
    \item[Auditory - Permanent] Regarding auditory cues, we first introduced a pair of sounds that deliver continuous cues regarding the temperature of the water. For an expected warmer sensation, we implemented a continuously looping $Boiling~Sound$. Because colder states of water generally do not emit continuous sounds, we selected a substitute association and integrated the sound of ice cubes that continuously move around colliding with each other and the glass ($Ice~Cubes~Sound$).
    \item[Auditory - Transient] Lastly, in our everyday lives, sounds are most often generated by the sudden change in temperature when touching an object. Therefore, we implemented a transient sound synchronized to the touch of the user. We play a $Sizzling~Sound$ akin to a water droplet hitting a hot surface and the sound of ice cracking when touched with warm hands ($Ice~Cracking~Sound$), which we expect will respectively increase and decrease perceived temperature.
\end{description}

\subsection{Apparatus}
\label{Apparatus}
To be able to generate actual changes in temperature for comparisons, we utilize Peltier elements. For this, we constructed five modules, each consisting of a 3D-printed housing, two Peltier elements, two attached temperature sensors, a microcontroller, and wiring and converters for the power supply. We affixed the two Peltier elements ($40mm\times40mm$) for each module vertically to opposite sides of the 3D-printed mount so that they can be touched with the thumb and index finger, respectively. The entire module's width measures $ 6 cm$ to allow for a comfortable grasp, and the Peltier elements' center was placed $ 5.5 cm$ above the base to avoid collisions. 
We attached temperature sensors\footnote{KY-001 temperature sensor from Joy-IT: \url{https://sensorkit.joy-it.net/en/sensors/ky-001}, accessed: 2025-07-11} directly to the Peltier elements with thermal paste for accurate measurements. 
We manually tuned proportional–integral–derivative (PID) controllers running on the microcontrollers to set each Peltier element to a desired, constant temperature. We then repeated this setup for all five modules and subsequently mounted them on a large wooden board. Each module was spaced $ 23.5 cm$ apart and turned $45$ degrees counterclockwise to allow a more comfortable interaction with the right hand.
To be able to change which module is positioned in front of participants, we placed the entire setup on a rail with a programmable linear actuator ($ 148 cm$ length) and calibrated it to the five module positions. The main components are illustrated in \RefApparatus{A}, and the working setup is displayed in \RefApparatus{B\&C}.

The visual and auditory stimuli (see \autoref{Stimuli}) were integrated into a virtual environment built in Unity3D\footnote{\url{https://unity.com/}, accessed: 2025-07-11} and rendered with its High-Definition Rendering Pipeline (HDRP)\footnote{\url{https://unity.com/features/srp/high-definition-render-pipeline}, accessed: 2025-07-11}. The virtual scene was displayed at 90Hz on an HTC Vive Pro head-mounted display (HMD) connected to a desktop computer\footnote{Intel Core i7-6700K processor, 16GB RAM, NVIDIA GeForce GTX 1080 graphics card}. We attached an Ultraleap Leap Motion Controller\footnote{\url{https://www.ultraleap.com/}, accessed: 2025-07-11} to the front of the HMD to track participants' hand and finger movements at 120Hz. Their right hand was represented in the virtual environment using a common low-poly model provided by Ultraleap. 
The virtual object is displayed in front of the participants as a cylindrical glass of water with the same diameter as the physical modules ($ 6 cm$). We chose a cylinder to avoid potential clipping through edges while participants grasp the virtual object. The physical and virtual environments are synchronized using a VIVE Tracker. The virtual view is presented in \RefApparatus{D}.

\begin{figure*}[t]
    \centering
    \includegraphics[width=\linewidth]{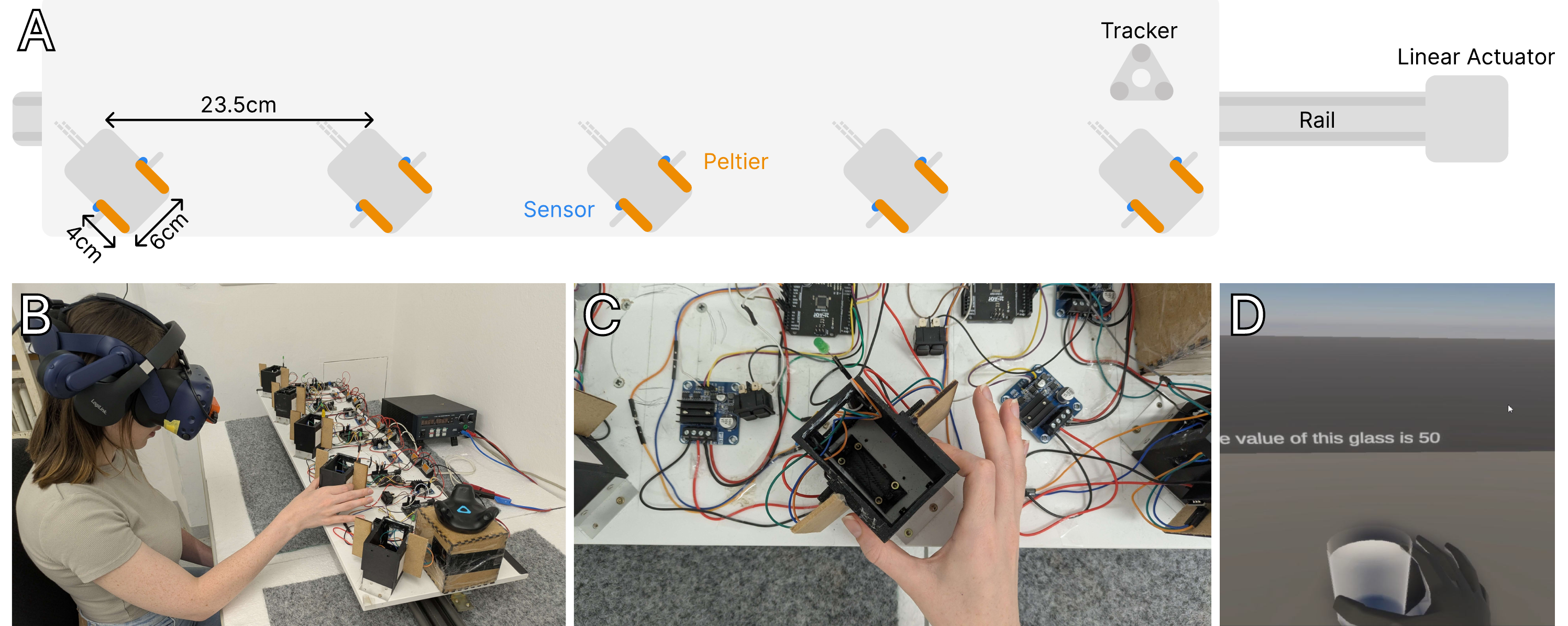}
    \caption{ (A) Our apparatus comprises five closed-loop modules spaced apart on a board, which can be moved back-and-forth along a rail. 
    (B) Participants sat in front of the apparatus and touched the modules while immersed in VR.    
    (C) Each module consists of two Peltier elements and two attached temperature sensors. The Peltiers are interacted with using the index finger and thumb of the right hand. (D) In VR, the object is visually displayed as a glass co-located with the module, and the participants' hands are represented by a low-poly model. 
    }
    \label{fig:apparatus}
    \Description{This figure is split into four parts (A-D). (A) shows an illustration of the apparatus. It shows five modules placed on a board, each with two Peltiers and two sensors, a tracker, and a rail with a linear actuator that moves the board forward and backward. (B) shows the same apparatus as an actual image, with a participant grasping one module while in VR. (C) shows a close-up of the grasp, where the index finger and thumb are placed on the Peltier modules. (D) shows the virtual view with a virtual glass with grey liquid and a low-poly hand representation.}
\end{figure*}

\subsection{User Study I: Subjective Magnitude of Temperature Change}
\label{Study1}
The aim of the first user study was to systematically scale the subjective temperature changes of the investigated stimuli to be able to order and compare their subjective effects across participants with each other. 
For this, we employed an established psychophysical Magnitude Estimation procedure~\cite{Jones2013_PsychophysicsInHaptics, Gescheider1997_Psychophysics}: For each trial, participants were tasked to first touch a reference object (baseline) and then the manipulated object (with visual or auditory illusions applied). They then rated the latter in relation to the reference by giving it an abstract value without restrictions. In addition to the visual and auditory stimuli (\IVIllusion) presented in \autoref{Stimuli}, we varied the physical temperature (\IVTemp{}) between trials.  We used three levels of physical temperatures -- 30\celsius{}, 35\celsius{}, and 40\celsius{} -- which are within the safe, comfortable range~\cite{Jones2008_MotivationPhysTemp, Gerrett2016_MotivationPhysTemp} and we determined in pilot tests to be noticeably distinct. These enable the measurement of the efficacy of \IVIllusion{} dependent on the base temperature of the objects and allowed us to obfuscate the fact that physical temperatures did not change between the reference and test object.

Each combination of \IVIllusion{} (including the baseline stimulus) and \IVTemp{} was repeated 4 times, resulting in a total of 132 trials (11 \IVIllusion{} $\times$ 3 \IVTemp{} $\times$ 4 Repetitions). We randomized the trial order for each participant. For each trial, we recorded the subjective magnitudes assigned by participants.

\subsubsection{Task}
We tasked participants to compare their subjective estimations of two objects presented in sequence.
During the first interaction, participants saw the baseline condition as a reference, without any added visual or auditory illusions. The virtual object was colocated with the physical module, rendering a predefined temperature. Participants were tasked to touch the object with their index finger and thumb of their right hand for 3 seconds, after which they were prompted with visual instructions to release their grasp. 
One second after release, the second object is displayed. For this step, one of the 11 visual or auditory stimuli was applied to the object. Participants again were instructed to touch and then release the object after 3 seconds. 
After completion, a virtual keypad was displayed in front of participants. 
Participants were told that the first reference object was predefined to have an abstract value of 50. They were then asked to rate the second object's value in relation to the reference by typing in an integer number on the keypad. The keypad was chosen to allow for the input of numbers without any restrictions or boundaries to avoid framing effects. Participants interacted with the keypad using a virtual ray pointer from a Vive controller we put in their left hand.   After submitting a value, the next trial was started.
Physical temperatures were only varied between trials -- not within trials between reference and test -- to avoid impeding the magnitude estimation procedure. 

\subsubsection{Procedure}
First, we welcomed participants and informed them about the study's aim and data processing procedure. After giving written consent, participants were asked to fill out a demographics questionnaire and were then seated in front of the table holding the physical setup. Participants were informed about their task and put on the HMD. Then, participants initially ran through a training session consisting of 3 sample trials. After training, the 132 study trials were started. Participants took a break after 44 and 88 trials. Participants took on average 54 minutes (SD=10min) to complete all trials. 

\begin{figure*}[ht]
    \centering
    \includegraphics[width=\linewidth, trim={0, 30, 0, 0}, clip]{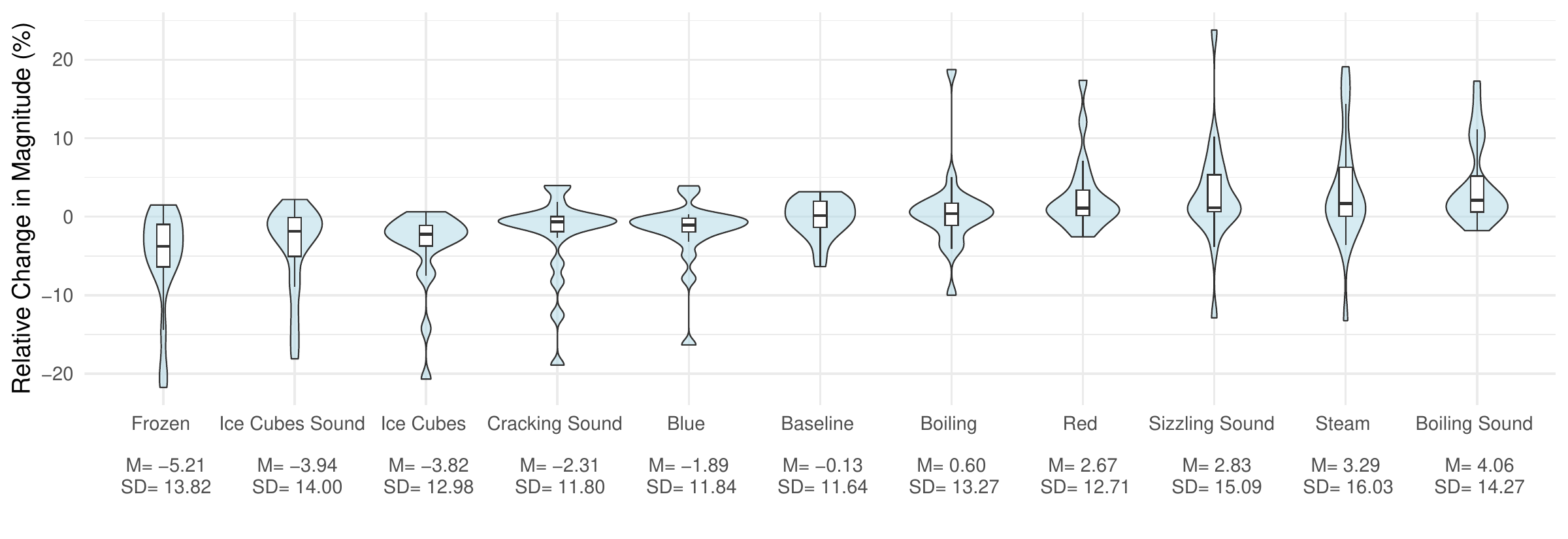}
    \caption{
    Distribution of normalized estimates for all levels of \IVIllusion{} averaged over all levels of \IVTemp{} and repetitions (ordered by mean). Displayed magnitudes are relative to the reference (ratio in \%) and normalized by participants' mean response.}
    \label{fig:magnitudeEstimations}
    \Description{This shows a box and violin plot for each stimulus level. We see that the visual and auditory stimuli affect the relative magnitudes compared to the reference. The order from lowest to highest judged temperature is: Frozen, Ice Cubes Sound, Ice Subes, Cracking Sound, Blue, Baseline, Boiling, Red, Sizzling Sound, Steam, Boiling Sound.}
\end{figure*}

\subsubsection{Participants}
We recruited 20 participants through university channels. 10 described themselves as female and 10 as male. Their age ranged from 17 to 35 ($M=24.90, SD=4.73$). 19 participants were right-handed and one person was left-handed. 18 had experience with VR (7 below 2h, 7 between 2h and 20h, and 4 for more than 20h). All participants had normal or corrected-to-normal vision and reported no issues regarding hearing or tactile perception. We offered each participant 15€ or university course credit as compensation. The study was approved by our institution's ethics board.

\subsubsection{Analysis}
To quantify the perceived magnitudes on a psychological scale, we followed established psychophysical procedures~\cite{Gescheider1997_Psychophysics, Jones2013_PsychophysicsInHaptics}. We first computed the relative ratio of reported responses compared to the reference. We then normalized every response per participant by dividing it by the grand mean of all of the participant's responses given during the experiment. This is necessary to eliminate individual variability in answers caused by participants being allowed to freely choose values without restrictions. We then computed the resulting psychological scale value for each condition and participant using the geometric mean~(\(\left( \prod_{i=1}^{n} x_i \right)^{1/n}\)) of all repetitions. 


For significance testing, we fitted Generalized Linear Mixed Models (GLMMs) using Laplace approximation implemented in the lme4 R-package~\cite{lme4}. We performed model selection based on Akaike (AIC) and Bayesian (BIC) information criteria. We chose a Gaussian distribution with log-link function with \IVIllusion{} and \IVTemp{} (without interactions) as fixed effects, and the participants' unique IDs as a random factor. 
We performed likelihood ratio tests (LRTs) comparing this model to reduced models, in which we individually dropped one fixed effect. Where we found a significant main effect, we conducted pair-wise post-hoc comparisons with Bonferroni correction. 

\subsubsection{Results}

The average relative changes in magnitude range from -11.9\% ($SD=14.1\%$) for the visual \textit{Frozen} condition at 30\celsius{} to +14\% ($SD=16.2\%$) for the visual \textit{Steam} condition at 40\celsius{}. We display the distributions, means (M), and standard deviations (SD) of relative perceived temperature change for each \IVIllusion{} averaged over all temperature levels and repetitions in \autoref{fig:magnitudeEstimations}. 
LRTs revealed a significant main effect of \IVTemp{} ($\chi^2(2) = 153.98, p <.001$) and of \IVIllusion{} ($\chi^2(10) = 44.83, p <.001$) on perceived temperature.
Post-hoc tests show significant ($p<.05$) differences between all pairs of \IVTemp{}, with 30\celsius{} ($M=-6.86\%, SD=12.76\%$) having the lowest estimate, followed by 35\celsius{} ($M=-1.96\%, SD=7.72\%$) and 40\celsius{} ($M=+7.77\%, SD=15.27\%$) with the highest perceived temperature.

For \IVIllusion{}, we found a significantly ($p<.05$) lower perceived temperature for the \textit{Frozen} visualization compared to \textit{Boiling Sound}, \textit{Steam}, \textit{Sizzle Sound}, and the \textit{Red} visualization. Similarly, the \textit{Ice Cubes Sound} had a significantly ($p<.05$) lower perceived temperature compared to the \textit{Boiling Sound} and \textit{Steam}. Lastly, the visual \textit{Ice Cubes} showed a significant ($p<.05$) decrease compared to the \textit{Boiling Sound}. 


\subsection{User Study II:  Quantifying Perceived Temperature Change}
\label{Study2}
After determining the subjective magnitudes of the illusions' effects on perceived temperature, we conducted a second user study to assess the size of these effects compared to physical temperature changes. This allows us to precisely quantify the influence of these illusions on perceived temperature (in \celsius{}). For this, we employed an established psychophysical discrimination task using the Method of Constant Stimuli with a two-alternative (two-interval) forced-choice paradigm~\cite{Jones2013_PsychophysicsInHaptics, Gescheider1997_Psychophysics}. We tasked participants to sequentially touch two objects and asked them to rate which one was perceived as warmer. One of the objects was always represented by a base temperature and a neutral visual stimulus. For the other object, we varied both the illusion application and the actual temperature presented by the physical modules. 

\begin{description}
    \item[\IVIllusion{}] To investigate both the individual and combined effects of visual and auditory stimuli, we selected stimulus pairs based on the results of the first experiment. Although the \textit{Frozen} visual and \textit{Boiling Sound} auditory cue produced the strongest subjective change, their cross-modal counterparts (\textit{Ice Cracking Sound} and \textit{Boiling} visualization) were considerably less effective. To enable a more balanced comparison and clearer interpretation of isolated and combined effects, we instead chose visual and auditory stimuli that independently produced similarly strong perceptual changes: \lvlSteam{} and \lvlSizzle{} for warmer, and \lvlIceCubes{} and \lvlIceCubesSound{} for colder sensations. 
    These were tested separately as well as their appropriate combinations, resulting in a total of 7 levels of \IVIllusion{} when including the neutral baseline (\baseline{}). We refer to these levels as warm visual (\warmV{}) and cold visual (\coldV{}), warm auditory (\warmA{}) and cold auditory (\coldA{}), and warm visuo-auditory  (\warmVA{}) and cold visuo-auditory (\coldVA{}) cues.
    \item[\IVTemp{}] In addition to the visuo-auditory presentation, we varied the physical temperatures. We used five levels of temperature: 28\celsius{}, 31\celsius{}, 34\celsius{}, 37\celsius{}, 40\celsius{}. The central value (34\celsius{}) is used as the base reference temperature. These values are comfortable and safe~\cite{Jones2008_MotivationPhysTemp, Gerrett2016_MotivationPhysTemp}, and distances between temperatures were determined by pilot studies to fit the requirements of the psychometric procedure (see~\cite{Gescheider1997_Psychophysics}). 
\end{description}

Each combination of \IVIllusion{} and \IVTemp{} was repeated twice, resulting in 70 trials per participant (7 \IVIllusion{} $\times$ 5 \IVTemp{} $\times$ 2 Repetitions). We balanced the order of reference and test object among repetitions and then randomized the trial order for each participant.
For each trial, we recorded the participant's binary decision. 

\begin{figure*}[ht]
    \centering
    \includegraphics[width=\linewidth, trim = {0 10 0 0}, clip]{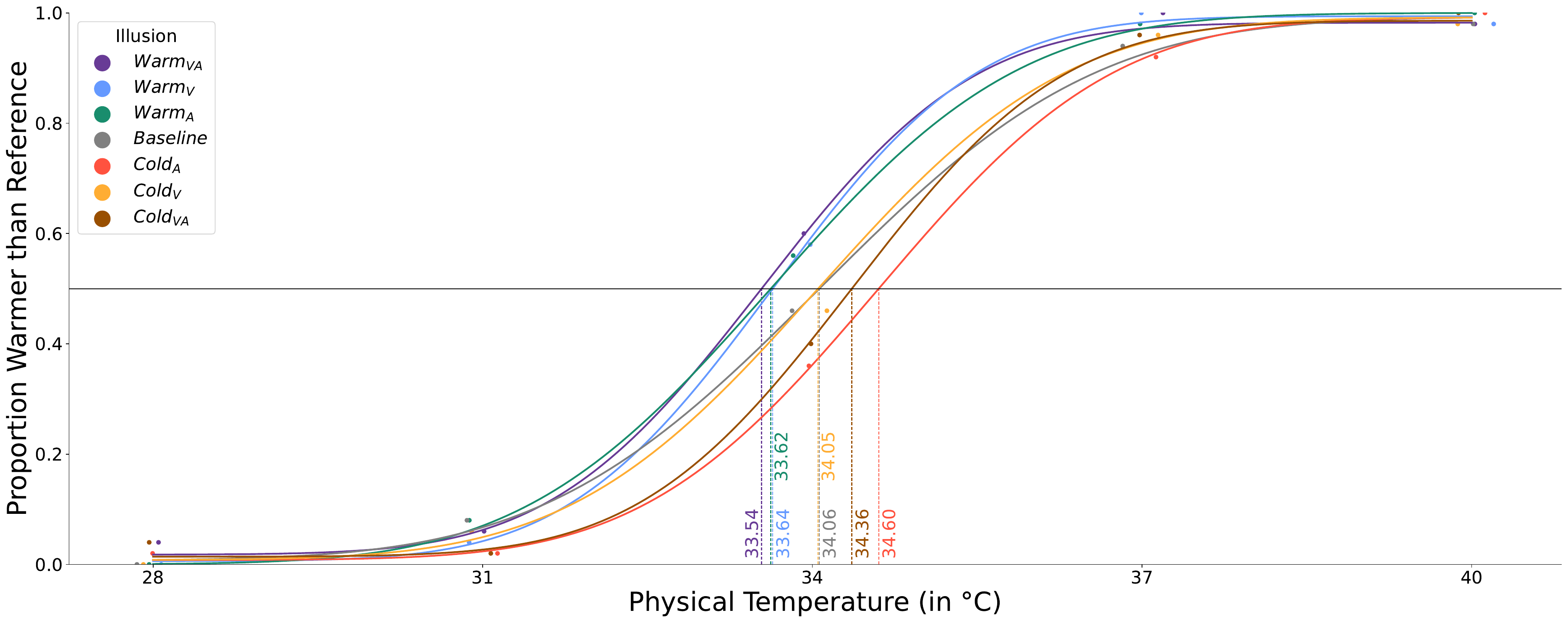}
    \caption{
    Fitted psychometric functions for each level of \IVIllusion{}. PSEs are defined as the points where each function crosses a proportion of 0.5, which is represented by a black horizontal line. Individual PSEs are indicated by vertical lines and labeled.
    }
    \Description{This shows seven psychometric curves with a sigmoidal form representing the different stimulus levels. We see the curves are shifted, with the warmer stimuli generally further to the left, and the colder stimuli more toward the right side. The points of subjective equality (50\% thresholds) are marked for each stimulus.}
    \label{fig:psyfit}
\end{figure*}

\subsubsection{Task}
Using the same apparatus as in the first experiment, participants were again asked to touch two objects in sequence. They were asked to touch each object for 3 seconds using the index finger and thumb of their right hand. Between the first and second touch, we varied both the applied illusion and the physical temperature by moving the rail setup to another module. During this delay, white noise was playing on noise-canceling headphones. If there was no change in required rail position, we still automatically moved the rail forward and backward in the same amount of time to obfuscate potential inferences based on vibration cues coming from rail movements. After completing both touches, participants were visually prompted to answer \textit{'Which of the two objects felt warmer?'}. They were only allowed to answer either the first or the second object by pressing left or right on a wireless presenter they held in their left hand. No feedback regarding their answer's correctness was given.

\subsubsection{Procedure}
We first informed participants about the task, received their written consent, and asked them to fill in a demographics questionnaire. Subsequently, we seated them in front of the apparatus and helped them put on the HMD. They ran through 3 training trials to show the phenomena they are supposed to assess: We once only altered the physical temperature, once only the visual stimulus, and once a combination of both visual and physical changes. After finishing the training trials, participants ran through all 70 study trials with a small break after 35 trials. On average, participants took 31 minutes (SD=3min) to complete all trials.

\subsubsection{Participants}
For the second user study, we recruited 25 new participants through university mailing lists. 14 described themselves as female and 11 as male. They were between 20 and 74 years old ($M=28.76, SD=12.05$). 21 were right-handed and 4 were left-handed. 18 had experienced VR before (6 below 2h, 7 between 2h and 20h, and 5 for more than 20h). All had normal or corrected-to-normal vision and no known conditions affecting hearing or the haptic or tactile perception of their hands.
They were offered either 15€ or a course credit as compensation. This study was approved by our institution's ethics board.

\subsubsection{Analysis}
We computed the ratio of the test object being perceived as warmer than the reference object for each combination of \IVIllusion{} and \IVTemp{}. For this, we divided the number of times the test object was selected as warmer by the total number of data points for each combination and averaged this across participants. 
We then fitted the resulting distributions for each \IVIllusion{} to a standard psychometric function~\cite{Gescheider1997_Psychophysics, Jones2013_PsychophysicsInHaptics}: We fitted
cumulative Gaussian distributions using the psignifit4 python toolbox for Bayesian psychometric function estimation~\cite{psignifit01, psignifit02}.

To assess the effect of the different illusions, we calculated the Point of Subjective Equality (PSE) for each \IVIllusion{}. These represent the points at which each test stimulus would be subjectively judged to be equal to the reference stimulus~\cite{Gescheider1997_Psychophysics, Jones2013_PsychophysicsInHaptics}. In our experiment, this corresponds to the point where the test object would be judged as warmer or colder an equal number of times, which occurs at a ratio of 50\%. 
To compute the effect of each illusion, we subtracted the PSE of the \baseline{} condition from the PSE of each illusion.  This effect corresponds to the actual physical temperature changes offset by the addition of the illusion, i.e., a physically warmer/colder temperature would feel equal to the baseline due to the illusion's effect.

Further, we calculated the just-noticeable difference (JND) of temperature -- i.e., the minimum necessary change in temperature necessary for successful distinction -- for each \IVIllusion{}. This is defined as the mean of the upper and lower difference thresholds, which are the distance between the PSE and the 75\% or 25\% points, respectively~\cite{Gescheider1997_Psychophysics, Jones2013_PsychophysicsInHaptics}. A lower JND indicates better acuity in distinction.

\subsubsection{Results}

The psychometric functions for the different levels of \IVIllusion{} are displayed in \autoref{fig:psyfit}. 
We observe horizontal shifts corresponding to the warmer and colder \IVIllusion{}. These represent the actual temperature (\IVTemp{}) that the illusions are offsetting.
Regarding the psychometric functions, the overdispersion parameter eta was consistently near zero ($<.001$) across illusion levels. This implies that variability in responses closely matched binomial expectations, supporting the robustness of the psychometric fits~\cite{psignifit02}.

The individual PSEs and corresponding illusion effects are presented in \autoref{tab:psyMetrics}. To improve interpretability, we inverted the sign of the effect so that positive values indicate a perceived increase (warmer) and negative values indicate a perceived decrease (colder) in temperature perception. 
The total range of the effect induced by the \IVIllusion{} was 1.066\celsius{}, from an increase of 0.527\celsius{} for \warmVA{} to a decrease of 0.539\celsius{} for the \coldA{} condition. 

JNDs generally surround a value of $\sim$1\celsius{}, with the lowest JND for \warmV{} (JND = 0.994\celsius{}) and the highest JND for the \baseline{} (JND = 1.337\celsius{}). We display the JND values of every \IVIllusion{} in \autoref{tab:psyMetrics}.


\begin{table}[t]
\centering
\begin{tabularx}{\linewidth}{r|XXX|XX}
\toprule
 Illusion & PSE & Upper  & Lower  & Effect & JND \\
\midrule
 \warmVA{}   & 33.537 & 34.556 & 32.519 & +0.527 & 1.019 \\
 \warmA{}    & 33.621 & 34.820 & 32.422 & +0.443 & 1.199 \\
 \warmV{}    & 33.638 & 34.632 & 32.644 & +0.426 & 0.994 \\
 \coldV{}    & 34.052 & 35.245 & 32.858 & +0.012 & 1.193 \\
 \baseline{} & 34.064 & 35.401 & 32.726 & 0.000  & 1.337 \\
 \coldVA{}   & 34.358 & 35.378 & 33.338 & -0.294 & 1.020 \\
 \coldA{}    & 34.603 & 35.737 & 33.470 & -0.539 & 1.133 \\
\bottomrule
\end{tabularx}
\caption{PSE (50\%), Upper (75\%) and Lower (25\%) difference thresholds, and resulting Effect and JND of each \IVIllusion{} in \celsius{}.}
\label{tab:psyMetrics}
\end{table}

\section{Discussion}
We conducted two experimental user studies to uncover and quantify the effects of visual and auditory presentations on perceived temperature. In the following, we discuss our findings regarding the efficacy of these illusions (\autoref{Disc:Efficacy}), the differences we found regarding subjective mappings and perceptual shifts (\autoref{Disc:ConsciousMapping}), the potential of auditory and multimodal cues (\autoref{Disc:Mutlimodal}), and what our findings implicate for the design and evaluation of future thermal VR experiences (\autoref{Disc:Implications}).

\subsection{Efficacy of Thermal Illusions} 
\label{Disc:Efficacy}
Prior work~\cite{Kocur2023_BurningHandsChangesBodyTemp, Villa2024_UltrasoundAndHueHeat, Guenther2020_Hydraulic_PlusHIVisual} has consistently shown significant differences in temperature judgments by participants while exposed to thermal illusions when using subjective, abstract scales. This indicates that thermal illusions are robust in altering perceived temperatures, but does not give concrete information about the strength of the effect. Our first study aligns with prior findings in that thermal illusions are judged to significantly alter perceived temperature. However, based on our second study, we found the strongest illusions to effectively shift perceived temperature in a range of $\sim$-0.5\celsius{} to $\sim$+0.5\celsius{}. This $\sim$1\celsius{} spread, from the \coldA{} to the \warmVA{} condition, represents the maximum perceptual change we observed, with other \IVIllusion{} variants producing subtler effects within these bounds. 
These effects are small compared to haptic illusion approaches targeting alternative haptic aspects. For instance, for stiffness perception, \citet{Weiss2023_VHI_Stiff} found visual manipulations in VR to induce up to $\sim$28\% softer and $\sim$8\% harder sensations. Similarly, \citet{Bergstroem2019_HandDistortion_Size} found physical objects could be virtually scaled up to 50\% larger sizes without user detection. Given the comparatively small changes introduced by the thermal illusions, they might be more suited for subtle feedback or to complement thermal rendering devices rather than serving as standalone mechanisms for strong temperature shifts.

Observed JNDs generally were much higher than in prior investigations focusing specifically on temperature acuity~\cite{Johnson1979_TemperatureAcuity}, which is unsurprising given the added noise of the virtual environment and task. Notably, we observed that the addition of thermal illusions did not negatively impact perceptual acuity, with grand average JNDs of each \IVIllusion{} condition lying below the JND of the \baseline{} visualization. This indicates that the discrepant information delivered to the visuo-auditory and haptic senses did not impact acuity beyond the factors also impacting the baseline, which would allow these thermal illusions to be utilized to enrich virtual environments without negatively affecting tasks relying on accurate haptic interactions.

\subsection{Influence of Conscious Mappings on Evaluations}
\label{Disc:ConsciousMapping}
While thermal rendering devices can be measured in objective metrics to allow comparisons, thermal illusions have to rely on subjective assessments to uncover potential changes in sensory perceptions. For this, it is often difficult to separate the subconscious change in perceived haptic stimulus intensity based on multisensory integration from the conscious mapping or ordering of clearly perceivable visual or auditory stimuli. For instance, one may be capable of ranking a set of color hues on a screen based on their temperature through conscious mappings without touching them. This phenomenon is separate from the perceived shift of temperatures induced by the integration of multiple senses during active haptic exploration. As opposed to some approaches below detectable thresholds from other areas of haptic illusions~\cite{Weiss2023_VHI_Stiff, Samad2019_VHI_PseudoHapticWeight, Bergstroem2019_HandDistortion_Size}, the visual and auditory stimuli applied during thermal illusions are always clearly noticeable~\cite{Guenther2020_Hydraulic_PlusHIVisual, Villa2024_UltrasoundAndHueHeat, Blaga2020_HI_VisualPresentations}, thus making subjective assessment generally weak towards large confounding effects caused by conscious mappings. We observe evidence of this during our user studies. Magnitude estimation, which produces ratios solely based on differences of the targeted stimulus, can be largely affected by conscious decision making. In contrast, our second approach using constant stimuli and offsets in physical temperatures is, while not completely unaffected, more robust against these effects due to the necessity of integrating the visuo-auditory as well as the haptic stimulus changes for the forced comparison. 
Here we found that subjective magnitude estimates and actual effects in \celsius{} do not always align. For instance, the \coldV{} condition (= $Ice Cubes$), which was rated similar to \coldA{} on the abstract psychological scale (\autoref{Study1}), showed no effect when we explored its capability of offsetting physical temperature changes (\autoref{Study2}). These differences based on evaluation procedure highlight the potential issues with established approaches relying on single subjective scales without comparisons to physical changes, potentially capturing unwanted phenomena instead of alterations of perceived temperature. 

\subsection{Potential of Alternate or Combined Channels}
\label{Disc:Mutlimodal}
While prior work focuses largely on visual approaches, we found auditory cues (\warmA{}: +0.443\celsius{}, \coldA{}: -0.539\celsius{}) to perform similarly or better than the studied visual analogs when investigated as isolated channels. This highlights the potential of audio-thermal illusions as alternatives or complements to the common visual presentation approaches. 
Regarding the potential of multimodal approaches, we observe that cues to separate channels, which already show clear effects when applied isolated (\warmV{} and \warmA{}), have a larger impact on perceived temperatures when combined. However, this combined influence is sub-additive, i.e., less than the theoretical sum of the isolated effects. 
Further, combining the visual and auditory cues with differing directions of effects (or no effects) seems to result in a combined percept between the separated outcomes. This is observed with \coldA{} (effect = -0.539\celsius{}) and \coldV{} (effect $\approx$ 0\celsius{}), where their combination results in a percept lying between their separated effect values (\coldVA{} = -0.294\celsius{}). This aligns with multisensory integration theory~\cite{Ernst2004_MultisensoryIntegration, Ernst2002_MLE}, which generally describes the resulting percept of multiple channels with discrepant information to center somewhere between the expected outcomes relying on each of these senses in isolation. For visual and haptic integration, \citet{Ernst2002_MLE} stipulated that their integration follows a maximum likelihood estimator, with senses weighted based on their reciprocal variances. While it is unclear if this holds for thermal perception and more than visual and haptic channels, the general assumption of a combined percept relying to some extent on the information of each channel is supported by the discovered effects of multimodal cues in our second study.

\subsection{Implications for Designing Thermal Experiences in VR}
\label{Disc:Implications}
Our findings offer several practical insights for designing thermal illusions in VR. While visual and auditory cues can shift perceived temperature, the effects we found are modest, typically within $\pm$0.5\celsius{}. Designers should therefore treat these illusions as subtle enhancements rather than strong replacements for actual thermal feedback, at least until the actual shifts in perception have been assessed for the respective thermal illusion in the respective context.  
Auditory cues, often overlooked, performed as well or better than visual cues in some cases. This highlights sound design as a viable channel for inducing thermal illusions, especially when visual attention is limited. Multimodal combinations partially showed enhanced effects, but unsuitable individual cues can reduce effectiveness, suggesting that separate evaluations for multimodal approaches are essential when targeting specific perceptual outcomes. Importantly, these illusions did not degrade perceptual acuity, indicating they can be integrated without impairing haptic task performance. However, designers should be cautious with evaluation methods: subjective scales may be influenced by conscious associations, and should be complemented -- wherever possible --  with methods that assess actual perceptual shifts.

\subsection{Limitations \& Future Work}
This work provides a first step towards understanding the effect of thermal illusions on the perceived temperatures of touched virtual objects. However, the intricacies of multisensory temperature perception and designing thermal experiences in VR introduce many practical and conceptual considerations that cannot be fully explored in a single work. 

Our findings and contributions are limited to the specific subset of visual and auditory thermal illusions examined in this study. The stimulus design was guided by prior work (\autoref{Stimuli}), and we employed a magnitude estimation procedure (\autoref{Study1}) to enable the inclusion of a broader range of illusions than would be feasible with the more restrictive discrimination procedure (\autoref{Study2}). However, the thermal illusions we explored represent only a portion of the possible stimuli that can influence temperature perception. Further research is needed to assess whether our results generalize to other types of stimuli and experimental contexts.

Further, the aim of our discrimination procedure (\autoref{Study2}) was to capture the general average effects of the investigated illusions and their cross-modal interactions. To enable within-subject comparisons across all conditions, we limited the number of repetitions, and we conservatively selected physical temperature levels to accommodate potentially strong illusion effects. Based on the observed PSEs and JNDs, future studies could explore smaller temperature differences to improve the precision of perceptual metrics, even though substantial increases in effect size are unlikely. Alternatively, focusing on a single or a small number of illusions would permit more repetitions per condition, enabling detailed analysis of individual differences in perceptual sensitivity and illusion strength.

Finally, while psychophysical methods that yield population-level thresholds are a valuable starting point for designing thermal illusion experiences, real-world applications will demand context-specific and individualized assessments. However, the time-intensive nature of these procedures—along with the complexity of the apparatus needed to generate precise and consistent comparison stimuli—may limit the practicality of conducting such assessments across different use cases. To address this, further research is needed to explore alternative evaluation methods. Physiological measures, for instance, have shown promise in detecting visuo-haptic illusions~\cite{Weiss2025_EEG4HI, Feick2024_IndividualBoundaries} and could offer a more efficient assessment route. Additionally, establishing practical baselines, reference comparisons, or standardized toolkits (as seen, e.g., for hand redirection~\cite{Zenner2021_VRHandRedirectionToolkit}) could help streamline and simplify future evaluation processes.

\section{Conclusion}

Thermal illusions offer a promising, lightweight alternative to hardware-based thermal feedback in virtual environments. However, prior work has primarily evaluated their effects using subjective ratings on abstract scales, leaving the actual extent of perceptual change unclear. To address this gap, we conducted two user studies using psychophysical procedures to both rank and quantify the impact of visual, auditory, and visuo-auditory thermal illusions on perceived object temperature. 

Our results show that while thermal illusions reliably influence subjective judgments, their actual effects on perceived temperature are modest, typically within $\pm$0.5\celsius{}. Auditory cues, in particular, showed strong potential as an alternative or complement to visual techniques, and illusions did not impair perceptual acuity, supporting their integration into VR experiences without degrading haptic task performance. However, we also observed clear discrepancies between subjective estimates and measured perceptual offsets, highlighting the limitations of evaluation methods based solely on abstract scaling. These findings emphasize the need for more rigorous, perception-based assessments. More broadly, the measured strength of the illusions calls for a rethinking of their role: rather than replacing physical thermal feedback, thermal illusions may be best used as subtle complementary enhancements or to allow differentiation without requiring strong temperature alterations.

\section{Open Science}
\label{OpenScience}
We provide access to our visual and auditory stimuli, collected datasets, data analysis scripts, and project files at this link: 
\url{https://osf.io/tx6bq/?view_only=5d5e3e0899784fd39a6a51c0ba2d5795}.



\bibliographystyle{ACM-Reference-Format}
\bibliography{bibliography}


\end{document}